\input psfig
\def\mathfont#1{\ifmmode{#1}\else{$#1$}\fi} %for math font     
\def\lae{\mathrel{<\kern-1.0em\lower0.9ex\hbox{$\sim$}}}  
\def\gae{\mathrel{>\kern-1.0em\lower0.9ex\hbox{$\sim$}}}  

\def\etal{{\rm et~al\/}.}

\def\und#1{$\underline{\smash{\hbox{#1}}}$}
\def\mcf{\mathfont{{\dot m}_{CF}}} 
\def\ergsec{\mathfont{ {\rm ergs\ s}^{-1}}}
\def\msun{\ifmmode{\ {\rm M}_\odot}\else{$ {\rm M}_\odot$}\fi}  
\def\msunyr{\ifmmode{\msun \ {\rm yr}^{-1}}\else{$\msun \ {\rm 
yr}^{-1}$}\fi}  
\documentstyle[11pt,paspconf]{article}

\begin{document}

\title{Optical Properties of Cooling Flow Central Cluster Ellipticals}
\author{Brian R. McNamara}
\affil{Harvard-Smithsonian Center for Astrophysics\\60 Garden St.\\
Cambridge, MA 02138}

\begin{abstract}
Central cluster galaxies in cooling flows show the signatures
of gaseous accretion and ongoing star formation at rates ranging
between $\sim 1-100\msunyr$.  Their
blue morphologies usually reflect the low net
angular momentum content of the $T<10^4$ K gas from which
the accretion population formed, and the
effects of interactions between the cool gas and their FR I radio sources.
For example, there is strong evidence that star formation is being triggered,
in part, by interactions between the $T<10^4$ K gas and the radio
sources in some objects.  Disk star formation on kiloparsec scales
is rare in cooling flows.  The optically determined star formation
rates, assuming the Local initial mass function (IMF), 
are typically factors of $10-100$ smaller than the cooling rates
determined from X-ray observations, and signatures of the
remaining material have not been identified outside of the 
X-ray band.  The IMF is poorly understood in
cooling flows;  most of the cooling material may be deposited in low-mass
stars or some other form of dark matter.  
Continued study of the interactions between 
radio sources and the intercluster medium will further 
our understanding of how elliptical galaxies, particularly
radio ellipticals in the early universe, evolve.

\end{abstract}


\section{Introduction}

Roughly half of the clusters that have been detected with X-ray
telescopes contain bright, centrally concentrated
X-ray emission from $\sim 10^7$ K gas that may be cooling from the hot phase
at rates of $\mcf \sim L_x T^{-1} \sim 10-1000 \msunyr$ (Fabian 1994).  
The cool material should settle
to the center of the cluster and accrete onto the 
central cluster elliptical galaxy, where it will collect
in molecular and neutral atomic gas clouds and eventually form stars.
Stars forming at rates comparable to the cooling rates
with a Local, disk-like IMF would render
the central cluster galaxies brighter and
bluer than normal ellipticals, and would be capable of producing
large centrally dominant elliptical galaxies (CDGs) if star formation continued
at rates of $\sim 100\msunyr$ over the lifetimes of the 
clusters, $\sim 10$ Gyr.  Optical, infrared, and radio spectral line 
surveys of CDGs in cooling flows do indeed indicate enhanced levels of
$<10^4$ K gas and star formation.  
However, the inferred star formation rates are much smaller
(e.g. $<1-10\%$) than the cooling rates, and the amounts of detected cold gas
can only account for $\lae 10^8$ yr of accumulated material.  
The discrepancy between the cooling rates determined
with X-ray observations and the detected accreted mass
is the so called ``cooling flow problem.''  Although the missing elements 
of this problem have not been identified conclusively, cluster cooling
flow CDGs often have 
remarkable properties that bear on several important issues in 
astrophysics, including the formation and evolution
of galaxies and radio sources in clusters, accretion dynamics in ellipticals, 
the nature of baryonic dark matter in clusters, and the suitability of CDGs as
standard candles.  This article
addresses the first three of these topics 
from the aspects of optical imaging and spectroscopy.

\section{Evidence for Accretion-Fueled Star Formation in Cooling Flows}

Several studies have shown that the spectral energy distributions
of CDGs in the wavelength range $\lambda \lambda
3600-5000$\AA~ exhibit continuum excesses and nebular line
emission whose strengths increase
with increasing \mcf\ (Hu \etal\ 1985; Johnstone et al. 1987; McNamara \& O'Connell
1989; Crawford \& Fabian 1993; Allen 1995; Cardiel et al. 1995).  
The color excesses have amplitudes in $U-B$ ranging 
from the detection limit of $\sim 0.1$ magnitude to $\gae 1$
magnitude, and the accretion populations contribute $\sim 10-70\%$
of the total light in the anomalously blue regions at U.   
Both the continuum excesses and the nebular emission
are often spatially extended over the inner 5-30 kiloparsecs of the CDGs
(McNamara \& O'Connell 1993; Heckman 1981; Cowie \etal\ 1983;
Heckman \etal\ 1989; Baum 1992; Donahue \& Voit, this conference).  
In addition, H I has been detected in absorption against
the radio continua of several CDGs in large cooling flows (O'Dea
\etal\ 1995; O'Dea, this conference), and molecular gas has been
detected in
the CO feature in Perseus (Lazareff \etal\ 1989) and 
in the $2.1\mu$ H$_2$ feature in essentially all of the 
large cooling flows with strong nebular line emission (Elston \&
Maloney 1996; Jaffe \& Bremer 1996).  These properties,
which are atypical of CDGs outside of cooling flows, 
indicate abnormally large levels of cold gas and star formation in 
the centers of cooling flow CDGs.  

$U-B$ color excesses in the central $\sim 5$ kiloparsecs
of CDGs in cooling flows, denoted $\delta (U-B)_{\rm nuc}$,
are plotted against 
the cooling rate of the hot gas derived from X-ray observations
in Figure 1.  The color excesses are plotted with respect to the mean 
for non-accreting template elliptical galaxies such that a negative 
excess indicates a bluer color.  The  optical data
are taken from the sources cited above, and the X-ray data
are mostly from Arnaud (1988).  The $U-B$ color was chosen because
it is commonly available and includes the U-band, which is
most sensitive to warm populations.  Some color excesses were
estimated from data in other bands and their errors  
may be $\gae 0.2$ magnitude, but most have errors $\lae 0.1$ magnitude.
The errors on the accretion rates are formally a factor of 2, but
systematic errors could be larger.  A trend for increasingly large blue 
color excesses with increasing cooling rate is seen in Figure 1. 
CDGs in clusters with $\mcf \lae 50\msunyr$ rarely show significant
anomalies.   M87's ($\mcf \sim 30 \msunyr$) blue synchrotron nucleus and 
jet is one exception.  CDGs in clusters with $\mcf \gae 200\msunyr$
show strong color excesses which often extend over the
inner 5-30 kiloparsecs of the CDGs.  The blue excesses in the moderate 
accretors ($\mcf \lae 200 \msunyr$) are often unresolved from the ground (e.g. A2052), and may 
be associated with weak AGN.  Dust features are present in at least half of
the galaxies observed (McNamara \& O'Connell 1992; 1993; McNamara
\etal\ 1996a; Pinkney \etal\ 1996).  This trend makes a strong case
for star formation fueled by accretion from cooling flows. Tidally
induced accretion, ram pressure accretion,
and mergers may affect this trend, although at first blush one would expect
these effects to erase the trend.  

\begin{figure}[h]
\hbox{
\hspace{.0in}
\psfig{figure=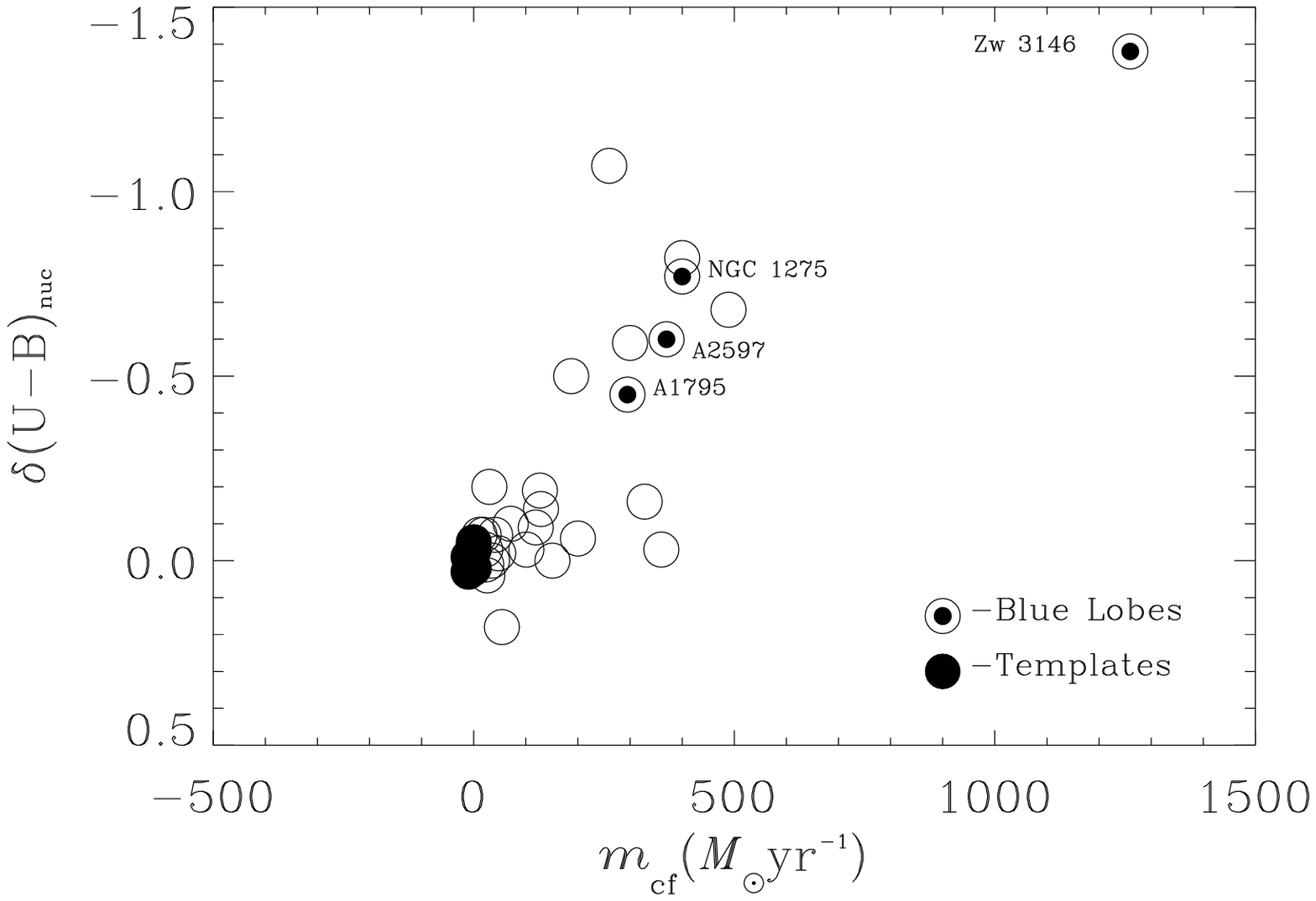,height=3.0in,width=3.5in}
}
\begin{minipage}[h]{5.0truein}
Figure 1: Correlation between central $U-B$ \und{continuum}
color excess and total cooling rate.
\end{minipage}
\end{figure}

\section{Star Formation Rates}

Although the correlation in Figure 1 can be viewed as a predictive
success of the cooling flow interpretation,  one's enthusiasm 
should be tempered by the theory's incompleteness.  While roughly half
of the gas accreted within
the extended blue regions (assuming $\dot m(r) \propto r$, see Fabian 1994) can be accounted
for, only a few percent or less of the {\it total} mass of
the gas cooling throughout the cooling region is found to be forming
stars with the Local IMF.  Furthermore, the optical properties of
some CDGs in clusters with large cooling rates are normal
(e.g. A2029).  The star formation rates are
determined in general as  $\dot S=M_{\rm lum} t^{-1}$.  The luminosity
mass is defined as $M_{\rm lum}= f_{\rm V} L_{\rm V}\Upsilon$, where
$f$ is the fraction of the total luminosity
in a given passband (V for example) contributed by the
accretion population, and $\Upsilon$ is the mass-to-light ratio
of the accretion population.  The age of the accretion population, $t$, is
in general poorly known and is usually taken to be the
typical age of a flow $\sim {\rm Gyr}$. $\Upsilon$ is taken from
stellar population models that match the assumed age and physical
conditions of the flow.  The star formation rates
determined in this manner are generally a few to a few tens of solar
masses per year (McNamara \& O'Connell 1989; 1993), compared
to accretion rates of a few tens to several hundred solar masses per
year.  These star formation rates
are in reasonably good agreement with those estimated using Balmer
emission (within a factor of 2), indicating
that the stellar continuum of the accretion population may be an 
important heat source for the nebular emission (McNamara \& O'Connell
1989).  This has been difficult to understand because emission line
spectra seem to be modeled better by shock ionization (Kent \&
Sargent 1979), irradiation from the condensing gas (Donahue \&
Voit 1991), or the turbulent mixing of hot and cold phase gas at the surfaces
of cold clouds (Crawford \& Fabian 1992), rather
than emission from classical H II regions.  In all likelihood
there are several heating mechanisms (Baum 1992; Fabian 1994), but
stellar photoionization is probably among the most significant 
(Donahue \& Voit, Cardiel, this conference).  

There is growing evidence, as I discuss below, that
a burst mode of star formation may be significant,
which would add uncertainty to the continuous rates.  In addition,
the presence of significant amounts of dust may lead to
underestimated star formation rates (Allen 1995).
The sink for the remaining accreted matter is unlikely to
be cold gas, as $<10^9 \msun$ of gas with $T<10^4$ K 
has been detected outside of the X-ray regime (Voit \& Donahue 1995;
O'Dea, this conference). 
The remaining mass may be deposited in low-mass stars forming with
an unusual IMF (Kroupa \& Gilmore 1994;  Schombert
\etal\ 1993; Sarazin \& O'Connell 1983) or some other form of dark
matter.  It is quite possible that the cooling rates have 
been overestimated to poorly understood gas physics.  However, 
cooling at levels of several to several tens of
solar masses per year in the largest accretors is consistent with
the optical data. {\sl AXAF} will provide important new 
observations with which to address this issue.

\section{Structure in Cooling Flow Central Cluster Galaxies}

The star formation histories of CDGs over the past several 
hundred Myr, and the dynamical state of the accreting
gas are imprinted on their surface brightness distributions and
color structure.
U-band imaging is most effective for studying the extended, blue
stellar populations and nuclear AGN light,  while being 
relatively insensitive to the
red background population.  The U-band's sensitivity
to blue continuum is somewhat limited by the [O II]$\lambda 3727$
emission feature, but becomes a less serious problem in galaxies at 
$z\gae 0.05$, as the emission marches out of the U passband in
the laboratory reference frame.  Galaxies are faint at
U, so 4m-class telescopes are required to study CDGs
in detail, and the number of well-imaged galaxies
is limited.  I will focus on what we have learned
from the best observed objects.

Cooling flow CDGs show a variety of blue structure.  Four
blue morphological types ordered by decreasing geometrical simplicity
can be broadly defined: 1. unresolved point source, 2. disk, 3. lobe,
4. irregular--amorphous.  These morphological types, shown schematically
in Figure 2, largely reflect the level of star formation, the angular  
momentum state of the accreting gas from which stars
are forming, and the physical relationships
between gaseous phases--the radio plasma in particular--and
star formation.  Type 1 galaxies contain unresolved    
blue nuclei and nebular line emission.  Their color excesses are 
modest, being one or two tenths of a magnitude in $\delta (U-B)$, and
they tend to occur in CDGs with modest accretion rates of $\lae 200 \msunyr$.  Their
nebular luminosities of $\sim 10^{41} \ergsec$ are factors of
several to ten smaller than those in the largest cooling flows (Heckman
\etal\ 1989), and
their color excesses signal either modest nuclear starbursts and/or
weak AGN. Typical objects in this class are the CDGs in A2052 and A2199.
Type 2 CDGs contain disks of gas and young
stars in rotation about the nucleus.
The archetype, Hydra A, contains a young, blue stellar population
that is confined to a $\simeq 10$ kiloparsec disk of neutral and ionized 
hydrogen and molecular gas.  The Type 2 
morphology reflects a high angular momentum
state of the accreted gas.  Type 2s are rare in cooling flows, but
not in the general population of radio galaxies (Tadhunter \etal\ 1989). 
Type 3, blue lobe CDGs are characterized
by bright, blue lobes of optical continuum located several kiloparsecs from 
the nucleus.  The best studied cases, A1795 and A2597, have radio 
emission that is coincident with their optical lobes (McNamara \& O'Connell
1993).  The  lobes are surrounded 
by nebular line emission with disordered velocities, suggesting
a low net angular momentum state (Heckman \etal\ 1989).  
Type 4, irregular--amorphous, encompasses most objects.
The gas velocities are again disordered 
(except perhaps close to the nucleus) reflecting a low net
angular momentum state.  As we shall see below, the blue lobes in Type
3s are probably
short lived and should evolve rapidly into an amorphous, Type 4 morphology.

\begin{figure}[h]
\hbox{
\hspace{.5in}
\psfig{figure=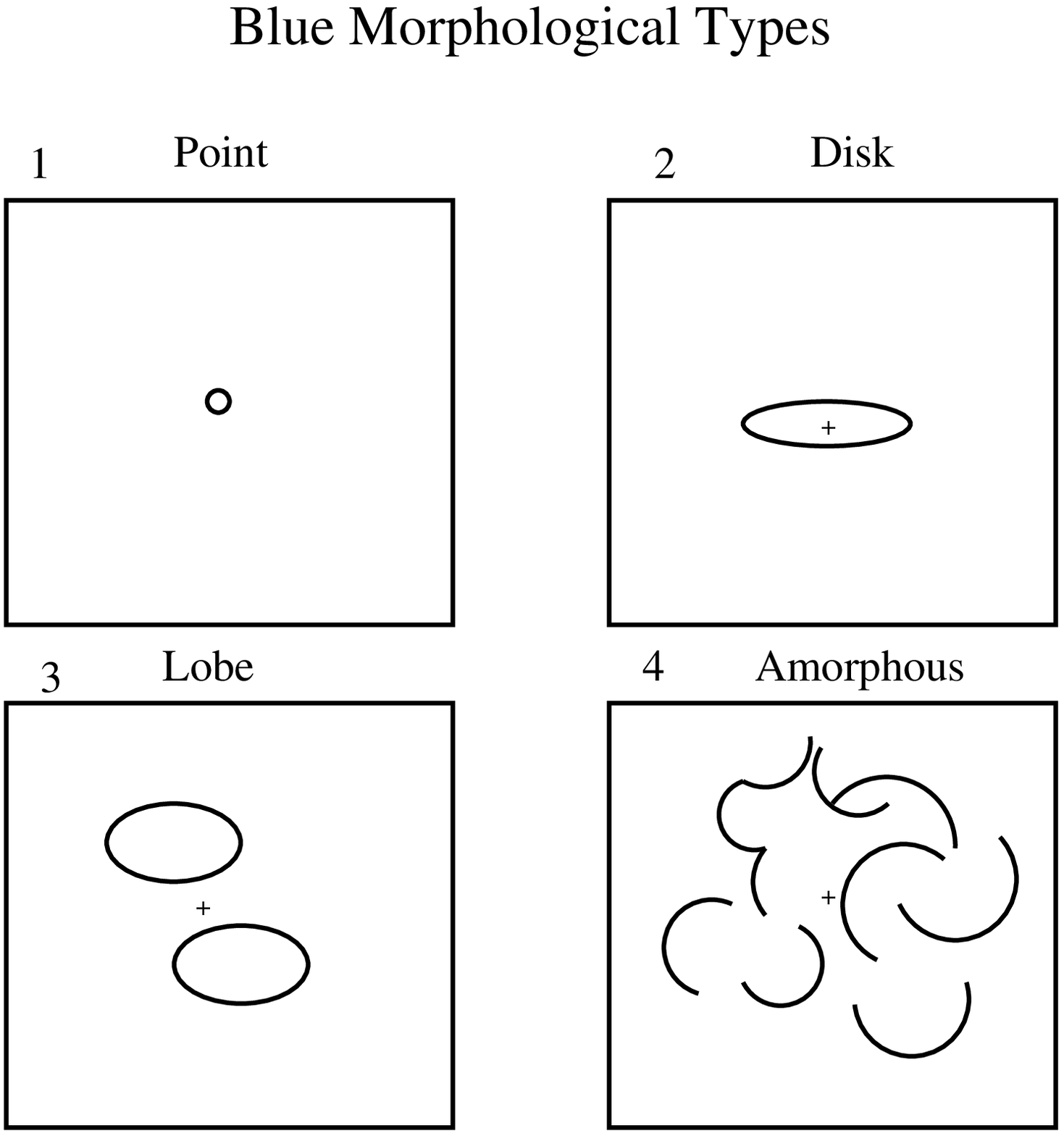,height=3.0in,width=3.5in}
}
\begin{minipage}[h]{5.0truein}
Figure 2: The blue morphological types found
in cooling flows. 
\end{minipage}
\end{figure}

\section{Interactions Between the Radio Source and Gaseous Medium}

The A1795 CDG, at redshift $z=0.06$, is
the archetype Type 3, blue lobe galaxy (Figure 3).  
A1795 and A2597 ($z=0.08$) harbor 
bright, blue lobes of optical continuum along the edges
of their radio lobes (McNamara \& O'Connell 1993; Sarazin \etal\ 1995;
McNamara \etal\ 1996a,c).  The lobes are surrounded 
by a diffuse, abnormally blue population.  A1795's FR I radio
source is $\sim 10$ kiloparsecs in size and its luminosity is
$L_{\rm radio} \simeq 9\times 10^{41} \ergsec$ (van Breugel \etal\ 1984).  
The radio jets emerge from the nucleus with a north-west, south-east
orientation and extend several kiloparsecs into the galaxy
before bending abruptly at right
angles and inflating into radio lobes at the location of the blue
optical lobes (Figure 3).  The jets were presumably deflected by
cold, dense clouds associated with the dust lane.
The optical lobes are resolved into discrete regions
along radio lobe boundaries in V and R HST images
(McNamara \etal\ 1996a; Pinkney \etal\ 1996). The
central 20--30 kiloparsec region is
embedded in luminous, $L({\rm H}\alpha)\sim 10^{42}\ergsec$, nebular
line emission (Heckman et al. 1989).  A2597 has similar
properties (Figure 4).

\begin{figure}[h]
\hbox{
\hspace{0.1in}
\psfig{figure=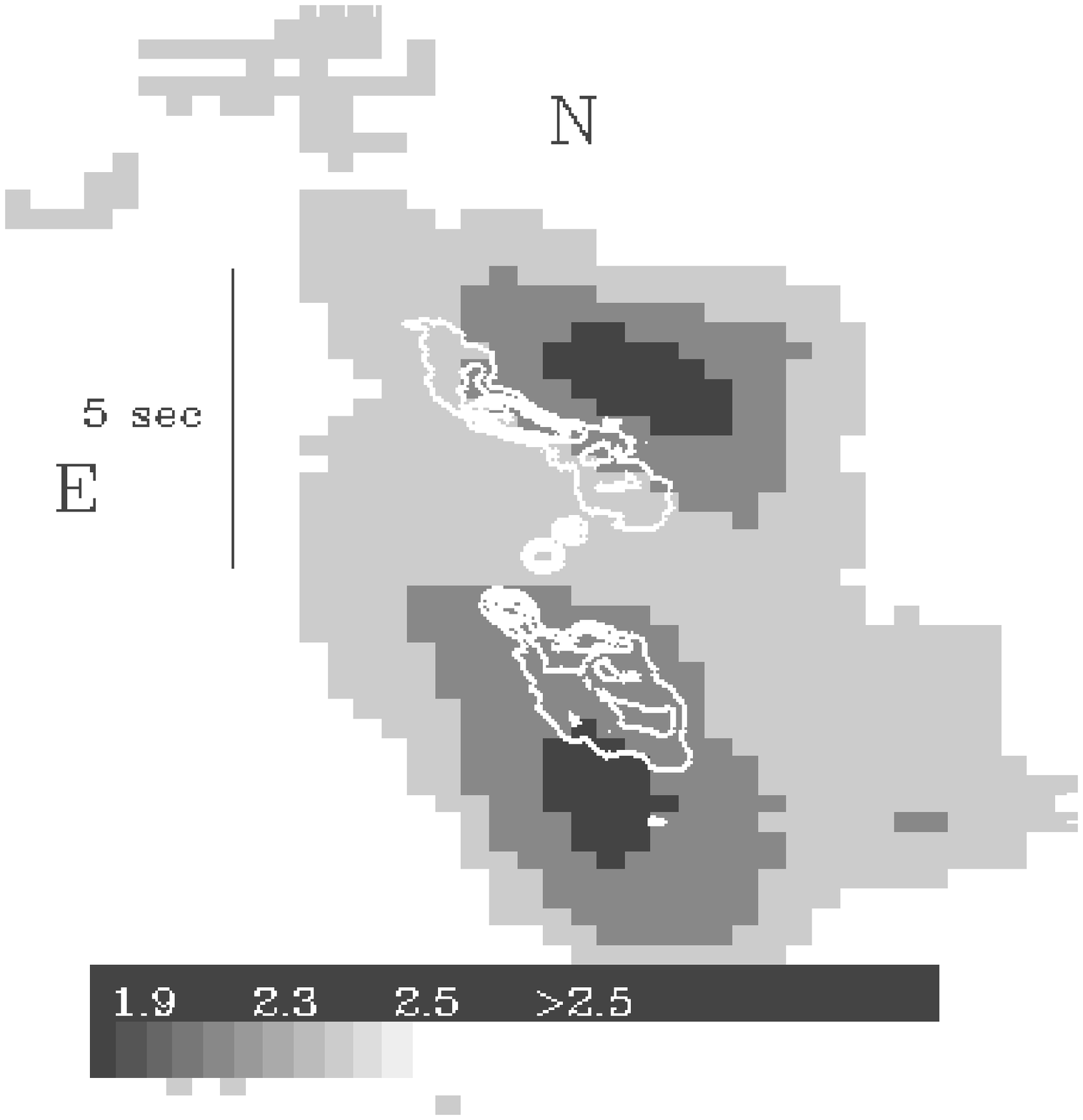,height=3.0in,width=3.in}
\psfig{figure=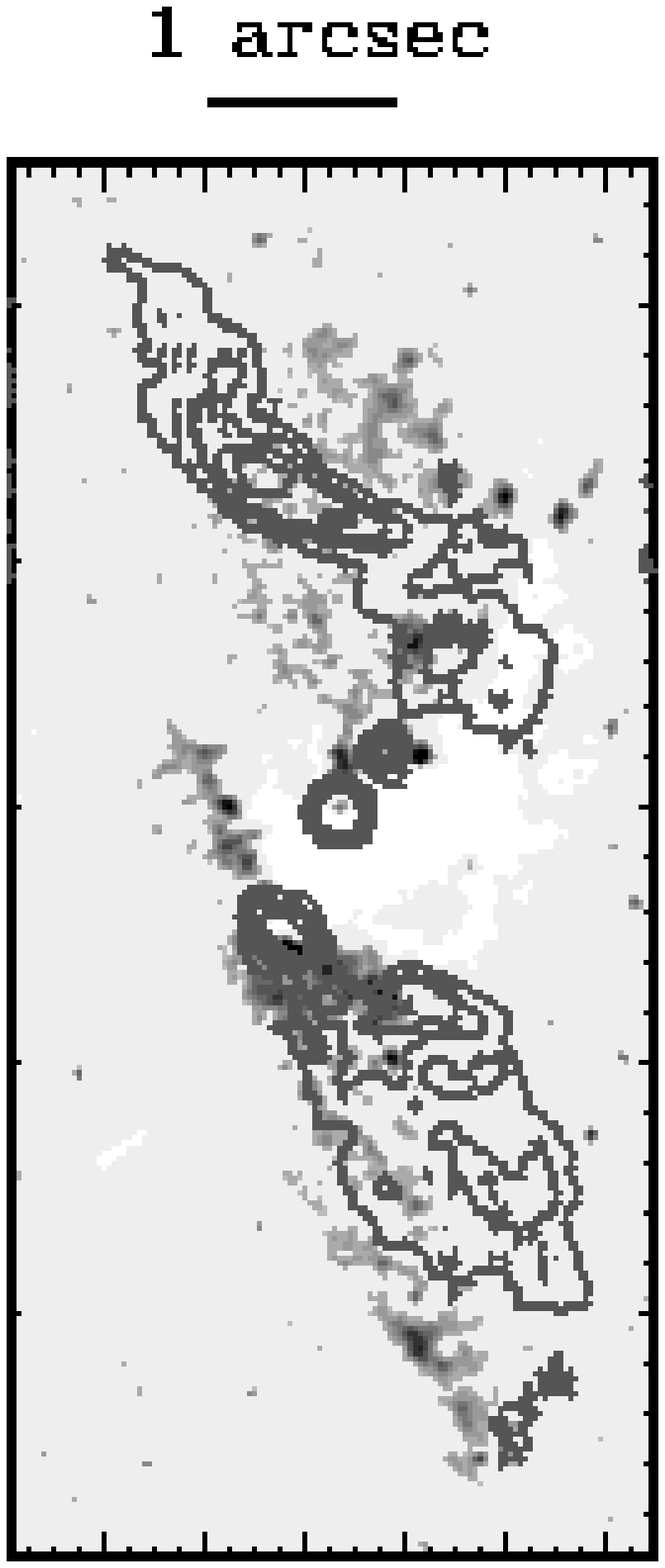,height=3.0in,width=1.4in}
}
\begin{minipage}[h]{5.0truein}
\vspace{0.1in}
Figure 3: (Left) Ground-based $U-I$ color map of A1795 showing the blue 
lobes (greyscale;
2 arcsec resolution) along the 3.6 cm radio lobes (contours; 0.2
arcsec resolution)[McNamara \& O'Connell 1993]. (Right) HST image
in V+R of the same region after subtracting a model
for the galactic background.  The white region along the jet is the
dust lane (McNamara \etal\ 1996a). 
\end{minipage}
\end{figure}

The discovery of the blue lobes in A1795 and A2597 
(McNamara \& O'Connell 1993) 
prompted several groups to suggest that the lobes were 
scattered light from a misdirected, anisotropically emitting
AGN (Sarazin \& Wise 1993; Crawford \& Fabian 1993; Murphy \& Chernoff
1993).  This model predicts that the lobe light should be highly
polarized.  However, a recent polarization measurement 
of A1795's lobes in the U-band resulted in an upper limit of less than
7\% to the degree of polarization, which essentially excluded the
scattering model (McNamara \etal\ 1996c). The absence of a detailed correspondence between the radio lobes 
and the optical lobes excludes synchrotron or inverse 
Compton scattering as the emission mechanism.  Finally, the
HST image (Figure 3) clearly shows what appear to be blue star 
clusters along the 
edges of the radio lobes, rendering this object perhaps the
most convincing case for radio-triggered star 
formation.  The probable age of the radio source in A1795
based on synchrotron losses is several to ten Myr 
(van Breugel \etal\ 1984). There
is no evidence for rotationally supported gas out of which
the lobes formed, so they should
fall into the center of the galaxy in less than the free-fall
timescale of $\sim 40$ Myr.
The close spatial correlation between the radio and optical
lobes shows that the stars have not fallen significant distances from
their birthsites, implying an age of less than $\sim 10$ Myr.
Recall earlier suggestions that star formation in cooling flows
may occur with unusual IMFs that are depleted in massive stars 
relative to low-mass stars,
or possibly truncated at the highest masses (see O'Connell \&
McNamara 1989 for a discussion).  
This dynamical constraint on the lobe population age should limit
the range of physically plausible stellar population models for
A1795's lobes.  A comparison between $\sim 10$ Myr old burst population
model colors and HST photometry of the knots along the
radio lobes in the U and UV, after correcting for internal extinction, should
be sensitive to the IMF, in particular to an upper mass truncation
of stars more massive than $\sim$ B5V.
%\clearpage
\begin{figure}[h]
\hbox{
\hspace{0.0in}
\psfig{figure=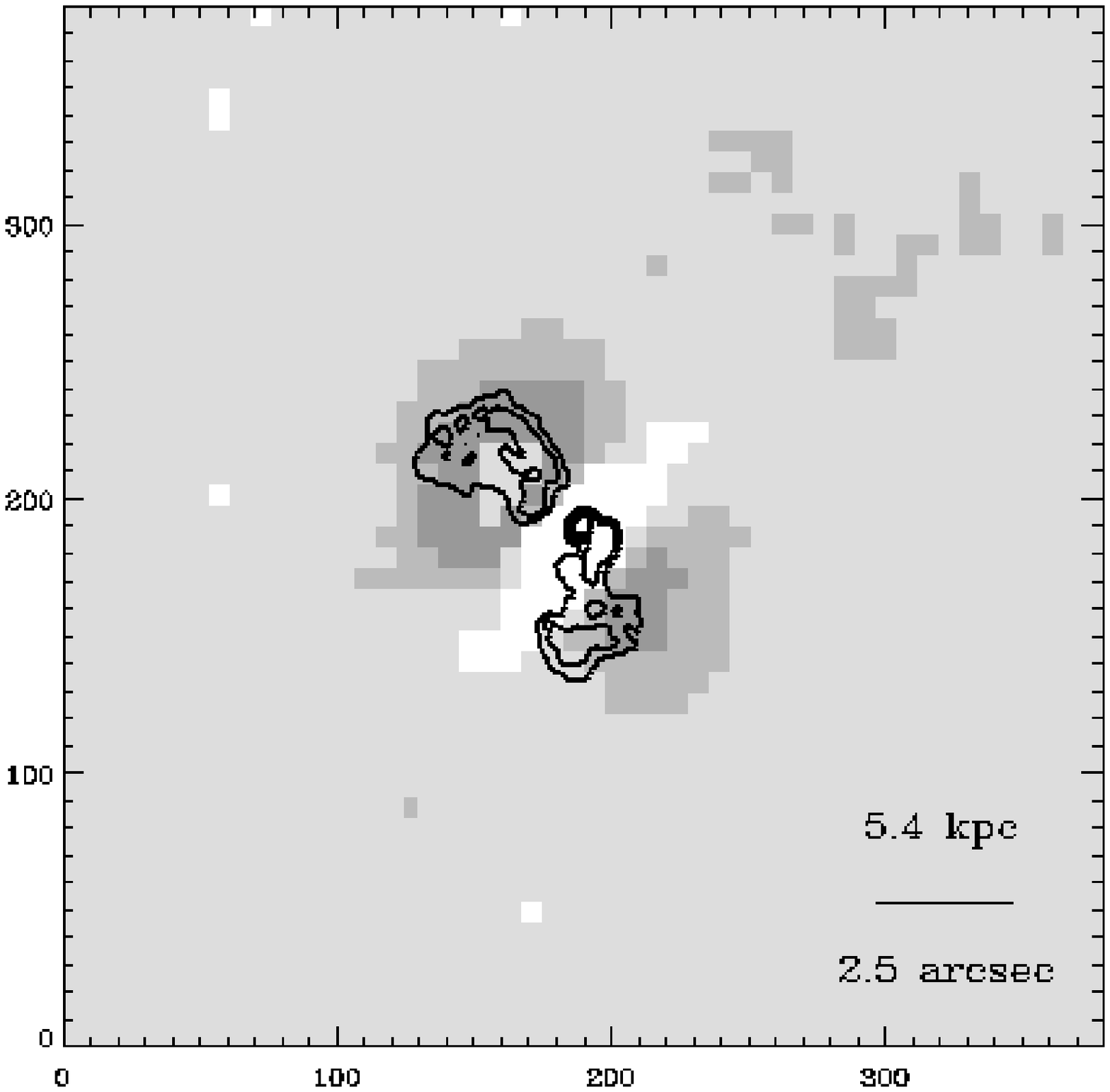,height=2.5in,width=2.5in}
\psfig{figure=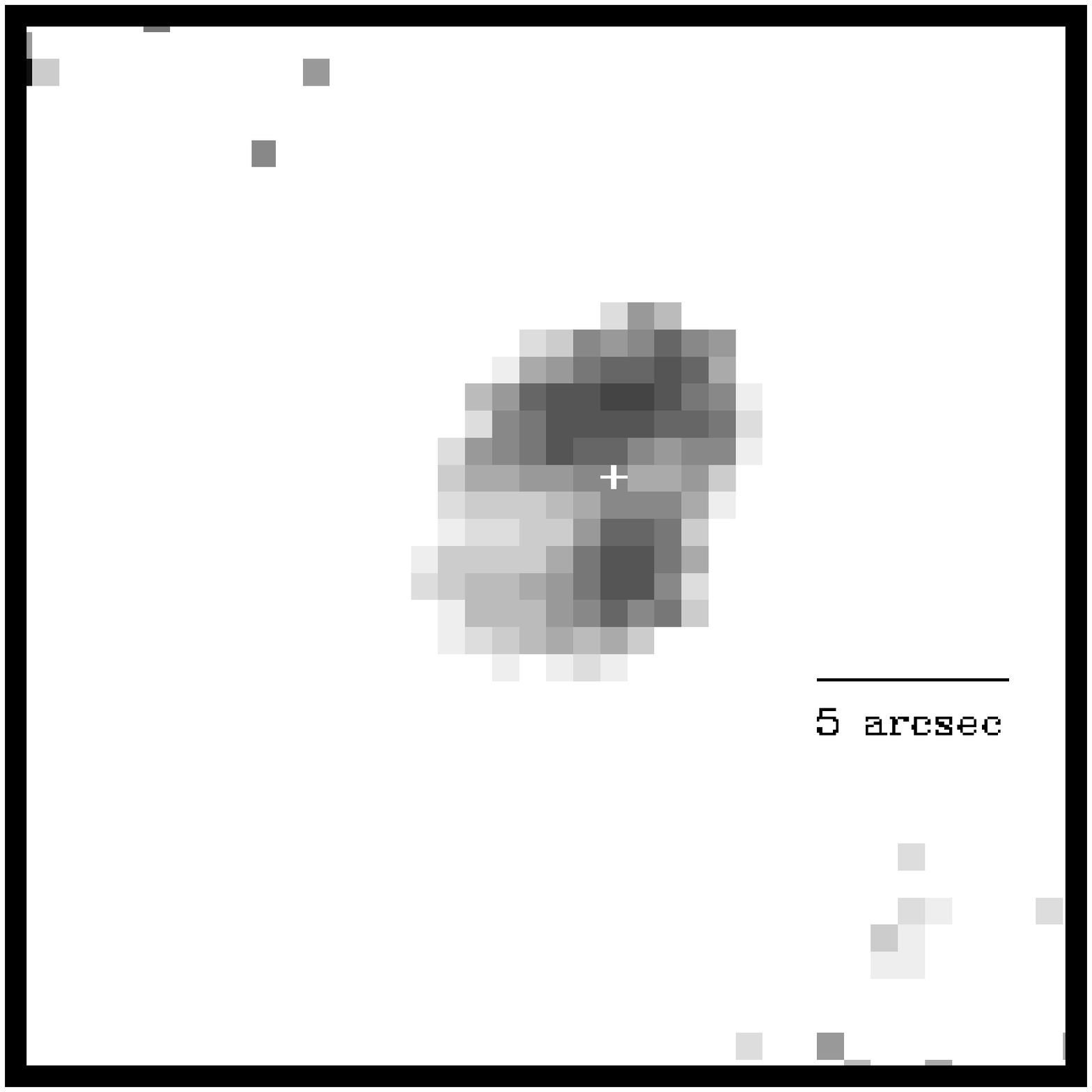,height=2.5in,width=2.5in}
}
\begin{minipage}[h]{5.0truein}
\vspace{0.1in}
Figure 4: (Left) Emission-line-free Str\"omgren b-band image of the 
lobes in
A2597 (greyscale) after subtracting a background model galaxy.
The 8.4 GHz radio map (contours) is superposed. A dust lane bisects
the nucleus at the location of the radio core.  The reversed greyscale
(darker for emission; lighter for absorption) is ``wrapped'' in the 
brightest part of the northern lobe for clarity (Sarazin \etal\ 1995;
McNamara \& O'Connell 1993).  (Right) $U-R$ color map of the 
distant, blue lobe galaxy in
the $z=0.29$ cluster Zw3146.  The restframe colors correspond to
mid-UV--V.  The intermediate passbands avoid
prominent emission lines.  The darkest regions
are $\sim 2$ magnitudes bluer than normal and are located a few arcsec
off the nucleus in a north-south orientation. The ``+'' indicates the location of the R-band nucleus. 
\end{minipage}
\end{figure}

Radio triggered star formation models usually attribute the triggering
mechanism to shock compression of cold clouds along the radio
jets (Rees 1989; De Young 1989; 1995). 
However, the star formation in A1795 was apparently triggered along the edges
of its radio lobes, rather than along
the jets, where shock compression should be most effective.  
The cold material may have collected along the lobes, creating a
gas overdensity that produced a locally enhanced region of star formation 
compared to the star formation that extends several kiloparsecs
beyond the radio lobes with a Type 4 morphology.
It is difficult to determine the fraction of the total star formation
rate associated with the lobes because the amorphous population
may have a different star formation history. Assuming the entire 
population formed
in a burst triggered less than 10 Myr ago, roughly half of the
$\sim 10^8 \msun$ of material involved in the burst could have been
induced to form stars by the radio source (McNamara \& O'Connell 1993).
It would appear that the radio source is augmenting existing
star formation, rather than triggering it outright.
The situation is probably similar in NGC 1275 (McNamara \etal\ 1996b) and perhaps in A2597, although polarization
measurements for these objects are not available.  A lobe-like
morphology (Figure 4) was recently discovered in the distant, $z=0.29$,
CDG Zw3146 (McNamara, Lotz \& Mackie 1997).
The lobes are not sufficiently resolved to
determine whether they are physically distinct or
are produced by a very large dust lane bisecting the nucleus.
In addition, high resolution radio maps are not available for this object,
whose radio luminosity is consistent with only an average
FR I, while its cooling rate exceeds $\sim 1000\msunyr$ (Allen \etal\ 1992).  
Based on its H$\alpha$ emission line luminosity, Zw3146's star
formation rate may approach $\sim 100 \msunyr$.
The lobe structure in Zw3146 
indicates that the blue-lobe phenomenon in cooling flows may
occur frequently over a significant fraction of cluster ages. 
A high resolution radio image of this object is badly needed.

\section{Disk Star Formation}

The Type 2 morphology differs from the remaining types
by the large angular momentum state of the accreted gas and stars.
Type 2s are rare in cooling flows, but they are important.
The Hydra A cluster central elliptical, which is the archetype
for Type 2s,  contains a blue, circumnuclear disk (Figure 5), 
$8 \times 6$ arcsec in size ($12 \times
9$ kiloparsecs; $z=0.055$, ${\rm H}_{\rm o}=50~{\rm km s}^{-1}{\rm Mpc}^{-1}$)
that is embedded in bright nebular emission in rotation about the
nucleus (McNamara 1995; Hansen \etal\ 1995; Heckman \etal\ 1989; Baum \etal\ 1988).  
The blue population is confined to
the disk, as is shown by the disk's 
exponentially declining light profile indicating ordered stellar orbits.
The Types 3 and 4 objects, in contrast, harbor a considerable
amount of diffuse blue light, presumably from young, infalling
stars on orbits with predominantly radial velocity components. 
Hydra A's most striking property is the orientation of the
radio jets nearly along the minor axis of the disk (see Figure 5). 
A twin-jet radio source is
thought to be aligned with the minor axis (angular momentum axis) of 
a subparsec accretion
disk surrounding a nuclear black hole (Rees 1984).  The apparently aligned angular momentum
axes of the putative subparsec and kiloparsec scale disks suggests 
they are coupled and may have formed simultaneously. 
Hydra A has long been a puzzle because its radio luminosity
($P_{178}=7\times 10^{43}\ergsec$) is an order of magnitude larger 
than most FR I radio sources (Ekers \& Simkin 1983), perhaps by the provenance of the disk.   
The 21 cm line of atomic hydrogen 
has been detected in absorption through the disk against the radio source 
(Dwarakanath \etal\ 1995).  
Roughly $\sim 10^{8\rightarrow 9} M_\odot$ of H I may
be present over the face of the disk.  Both the radio jet axis and
the disk's principal
axes are misaligned with the principal photometric axes of the
host galaxy, suggesting the gas was externally accreted.
The origin of disk's angular momentum is unknown.
It is noteworthy that star formation was apparently not triggered by
the radio source, in spite of Hydra A's large radio power and dense
intracluster medium (David \etal\ 1990), probably because the 
radio source has encountered little cold gas. 
In all likelihood, the nebular emission traces the cold gas which remains close
to the disk's orbital plane and away from the disk's polar regions
where the jets are emerging.  In contrast, the gas in Types 3 and 4 is 
widely distributed, and their radio sources have a high
probability of colliding with gas clouds, resulting in distorted radio
lobes, disrupted jets, and radio-triggered bursts of star formation.   

%\clearpage
\begin{figure}[h]
\hbox{
\hspace{1.0in}
\psfig{figure=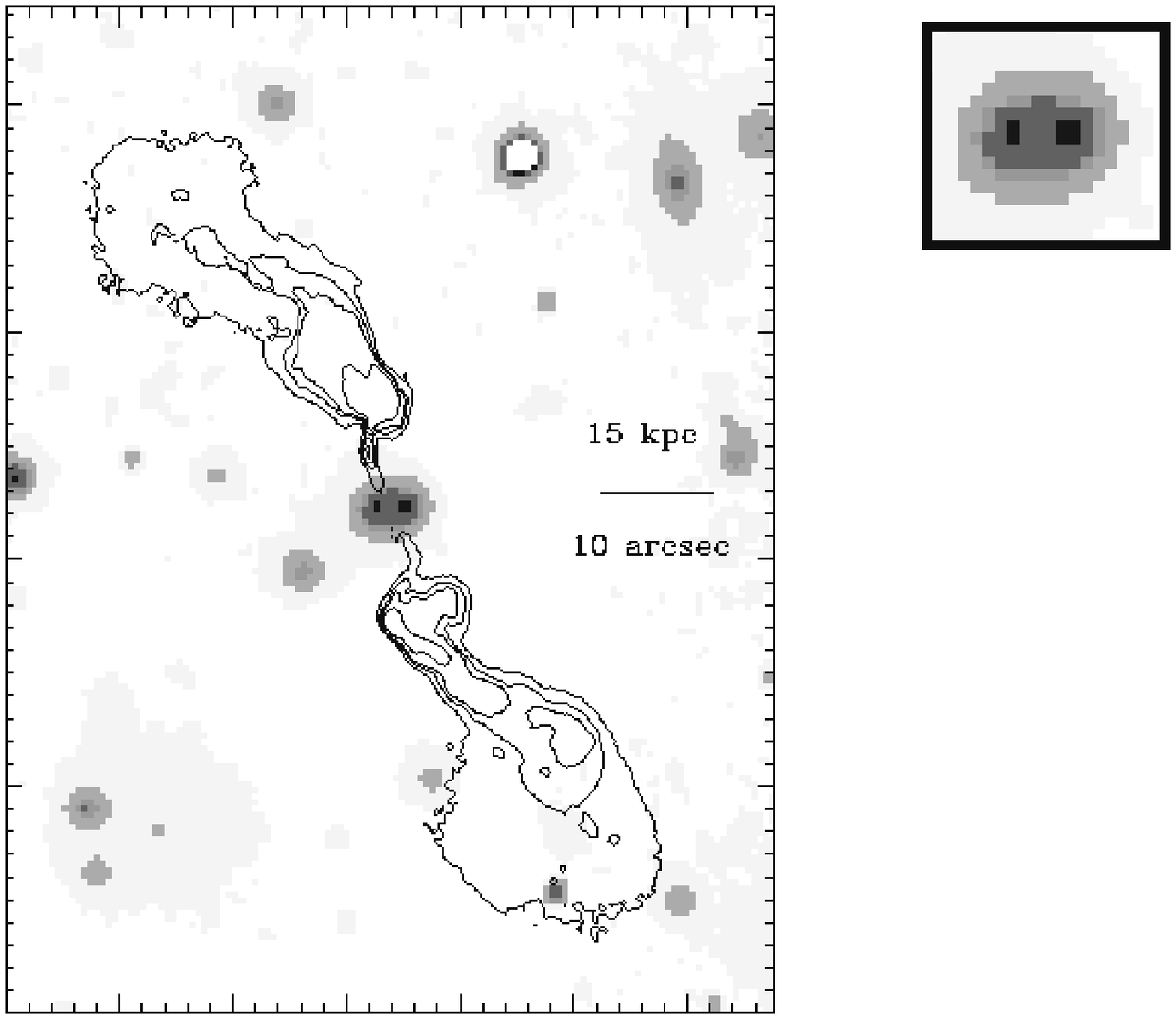,height=3.0in,width=3.in}
}
\begin{minipage}[h]{5.0truein}
\vspace{0.1in}
Figure 5: U-band map of the central disk (greyscale) in Hydra
A after subtracting a model galactic background. An expanded view of
the disk is inset to the upper right.  The 6 cm radio map (contours) 
is superposed; the  central radio core is omitted for clarity (McNamara 1995). 
\end{minipage}
\end{figure}

The ordered motion allows, in principle, 
a direct measurement of the total mass (disk + galaxy) encircled by the disk.
The mass ratio of stars in the disk and underlying host galaxy
can then be determined by modeling the background galaxy and
photometrically decomposing the galaxy and disk.
The dynamical and luminosity masses of the disk population 
can then be matched by adjusting $\Upsilon_{\rm disk}$, thereby
placing constraints on the IMF and dark mass.
The required observations are best done with HST, but an analysis
using ground-based
observations indicates a mass of the young disk population to be 
$\sim 10^{8 \rightarrow 9}\msun$, or $\sim 1\% \rightarrow 10\%$ of the total
encircled mass (McNamara 1995).  Existing ground based observations
cannot be modeled precisely enough to infer useful information on the IMF.

\section{An Evolutionary Sequence?}

There is enough data on the structure of CDGs that one can begin
to speculate on whether the structural types reflect
an evolutionary sequence.  In the simplest scenario, the CDG
accretes $\sim 10^{8}\msun$ of material from a
gas-rich dwarf galaxy or a cooling parcel of hot gas.  This
material sinks to the nucleus which generates
an FR I radio source, followed shortly by 
a radio-triggered burst of star formation fueled by the infalling gas.
The initial result is a Type 3.  The lobes quickly disperse in 
a few tens of Myr leading to a Type 4 morphology, while
its color decays
at the rate  ${\Delta(U-B) \over \Delta {\rm log} t} \simeq 0.45$
magnitudes dex$^{-1}$ (c.f. Bruzual \& Charlot 1993), 
and ends as a Type 1 in $\sim$ Gyr.  NGC 1275 in the Perseus cluster
is an example of a CDG that may be in transition between Types 3 \& 4 (see McNamara \etal\ 1996b).
This evolutionary
scenario may seem plausible based on the range of colors in Figure 1,
and while I think there are aspects of this sketch that are probably
true, it is too simple for several reasons.  First, the blue
lobe galaxies--Type 3s--are too prevalent.  Of the 20 or so
objects that have been imaged well enough to detect
lobes, 2--4 are Type 3's and there
may be more lost to projection.  The Type 3 phase should be
brief; it represents $< 1\%$ of the few Gyr lookback time of
the available sample in Figure 1.  If we are very lucky
we should see one Type 3, barring some unknown selection effect
(this sample was not selected on the basis of the radio properties).
Therefore, radio triggered bursts must occur several times during
the lifetimes of these galaxies, or they persist over long
timescales which would be difficult to understand.  For this to
be so, the fueling must be frequent or persistent.
There are other problems: Hydra A does not fit this scenario,
and the few objects that have been studied in some detail (e.g. NGC 1275)
seem to have stellar populations spanning a larger range of ages than
one would expect from a single burst (see McNamara \etal\ 1996b for
a recent discussion).  In addition, the apparently continuous distribution
of color excess when plotted against $\mcf$ suggests that the fueling is
not entirely random, as one would expect for tidally triggered
bursts, although some  scatter in Figure 1 could be explained by
periodic interruptions in the fuel supply by mergers, for example.
Perhaps the rare Type 2s, Hydra A for example, show the high angular
momentum signatures of tidal accretion, and perhaps not accretion
from the cooling flow.  Sensitive imaging at high resolution
(e.g. Pinkney \etal\ 1996; Holtzman \etal\
1996; McNamara \etal\ 1996a) of a large sample that includes a carefully
selected control subsample of CDGs outside of cooling flows
will be necessary to sort this out.

\section{Concluding Remarks}

It is clear that CDGs selected on the basis of residing
in large cooling flows have unusual optical properties, and these
properties are the result of gaseous accretion.
Whether the hot gas is providing the cooling material
that fuels the star formation, or whether the dense medium is promoting
accretion from other sources--e.g. by providing a large cross
section for ram pressure stripping of gaseous dwarf galaxies (McNamara
\etal\ 1996a)--remains to be conclusively demonstrated.  In my
view the strongest evidence points to cooling flow accretion,
but both processes are probably important (major mergers between
galaxies are probably less important).  The most vexing 
problem in cooling flows, the fate of most of the accreting mass,
remains unsolved.  I touched on two approaches for studying
the IMF that may lead to progress with this question, but other approaches,
such as the search for populations of low-mass stars in the infrared
may be fruitful.  We have not
identified conclusively a CDG that formed out of a cooling flow.
Zw3146 may be our best bet, but in general cooling flows are probably 
augmenting the mass of preexisting galaxies that grew, in part, by
mergers.  What these objects are telling
us about the evolution of ellipticals in general is most interesting.
The blue lobe galaxies are important in this regard,
because they may be related in some way to
the  distant, FR II radio galaxies showing the ``alignment effect.''
Many distant FR IIs have blue optical structure that is 
aligned with their radio
sources, and likewise may be undergoing massive bursts of star
formation  at rates of $\sim 100
\msunyr$ (McCarthy 1993).  The situation
is more complex in the FR II's, whose radio-aligned optical
continuum is often polarized (Jannuzi \& Elston 1991). 
The polarized continuum appears to be light from an AGN that is beamed along
the radio axis and scattered into the line of sight by dust or
electrons.  However, blue continuum from young stars 
may be a significant component (e.g. Stockton
\etal\ 1996).  A1795 may be a relatively nearby, lower luminosity, 
FR I counterpart that is
dominated by starlight rather than scattered AGN light. 
Clearly the alignment effect is not the exclusive domain of FR II
radio galaxies at high redshifts.  Further progress in this area
will be made using new, high spatial resolution imagery
and spectra of CDGs in the ultraviolet, U, and infrared bands,
and in the X-ray band with {\sl AXAF}.

\acknowledgments

I thank our hosts, Noam Soker and his colleagues in Haifa for
providing a stimulating atmosphere in which to discuss cooling flows,
and my colleagues with whom I have worked on this topic.  I thank
Bill Forman for helpful comments on the manuscript.   This research
was supported by grant NAS8-39073 to the Smithsonian Astrophysical Observatory.

\end{document}